\documentclass[%
 reprint,
superscriptaddress,
 amsmath,amssymb,
 aps,
pre,
]{revtex4-2}

\usepackage{graphicx}
\usepackage{dcolumn}
\usepackage{bm}

\begin{document}

\preprint{APS/123-QED}

\title{Chaotic Dynamics and Bifurcation Analysis of the Hindmarsh--Rose Neuron Model with Blue-Sky Catastrophe under Magnetic Field Influence}

\author{Ram Pravesh Yadav}
\email{ram1008pravesh@gmail.com}
\affiliation{School of Computational and Integrative Sciences, Jawaharlal Nehru University, New Delhi-110067, India}

\author{Hirdesh K.~Pharasi}
\email{hirdesh.pharasi@bmu.edu.in }
\affiliation{School of Engineering and Technology, BML Munjal University, Gurugram, Haryana-122413, India}

\author{R.~K.~Brojen Singh}
\email{brojen@jnu.ac.in }
\affiliation{School of Computational and Integrative Sciences, Jawaharlal Nehru University, New Delhi-110067, India}

\author{Anirban Chakraborti}
\email{anirban@jnu.ac.in}
\affiliation{School of Computational and Integrative Sciences, Jawaharlal Nehru University, New Delhi-110067, India}

\begin{abstract}
We investigate the impact of magnetic-field-induced feedback on the dynamics of a Hindmarsh--Rose neuron model exhibiting a blue-sky catastrophe. By introducing a magnetic flux variable that couples nonlinearly to the membrane potential, we demonstrate that electromagnetic effects profoundly reshape neuronal firing patterns and bifurcation structure. Interspike-interval bifurcation analysis reveals a nonmonotonic dependence on the magnetic coupling strength, with weak coupling preserving regular spiking and bursting, intermediate coupling promoting chaotic bursting, and strong coupling yielding structured irregular dynamics. These transitions are quantitatively characterized using the largest Lyapunov exponent computed via the Wolf algorithm and supported by Poincar\'e sections and time-series analysis. Our results establish electromagnetic feedback as a robust and tunable mechanism for controlling instability and chaos in slow--fast neuronal systems.
\end{abstract}

\maketitle

\section{Introduction}

Understanding the dynamical mechanisms underlying neuronal activity remains a central problem in computational neuroscience and nonlinear dynamics. Neurons exhibit a rich repertoire of electrical behaviors, including quiescence, tonic spiking, bursting, and chaotic firing, which are believed to play a crucial role in information processing and synchronization in neural systems. Mathematical neuron models provide an essential framework for uncovering the mechanisms responsible for such complex dynamics and for bridging microscopic ionic processes with macroscopic neuronal behavior.

Among reduced neuron models, the Hindmarsh--Rose (HR) model has emerged as a paradigmatic system for studying nonlinear neuronal dynamics due to its ability to reproduce both spiking and bursting activity using a minimal set of ordinary differential equations~\cite{HindmarshRose1982,HindmarshRose1984,RoseHindmarsh1989}. Owing to its low dimensionality and analytical tractability, the HR model has been extensively employed to investigate bifurcation structures, synchronization phenomena, and transitions to chaos in single neurons and neuronal networks~\cite{Rosenblum1996,Dhamala2004,GuPan2010,VinodGopal2012}. Detailed bifurcation analyses have revealed that the HR model exhibits multiple dynamical regimes organized by Hopf bifurcations, period-doubling cascades, and homoclinic orbits, making it a valuable testbed for nonlinear dynamical theory~\cite{Storace2008}.

Recent studies have emphasized that neuronal dynamics can be profoundly influenced by external electromagnetic effects, such as magnetic flux and induced electric fields, which naturally arise in biological tissue and can be externally applied through stimulation techniques. Incorporating electromagnetic coupling into neuron models introduces additional nonlinear feedback that can significantly modify firing patterns and stability properties. In this context, several modified neuron models have been proposed to explore the effects of electromagnetic radiation, memristive coupling, and non-autonomous forcing on neuronal activity~\cite{LvMa2016,ThottilIgnatius2019,Bao2019}. These investigations have demonstrated the emergence of multistability, mixed-mode oscillations, and novel bursting scenarios that are absent in classical formulations.

A particularly intriguing mechanism leading to sudden qualitative changes in neuronal activity is the so-called blue-sky catastrophe, a global bifurcation in which a stable periodic orbit disappears as its period diverges while its amplitude remains finite. In neuronal systems, blue-sky catastrophes have been shown to mediate abrupt transitions between spiking and bursting states and to generate complex firing patterns under stochastic perturbations~\cite{Bashkirtseva2019}. Despite its significance, the interplay between blue-sky catastrophe, chaos, and external electromagnetic effects in neuron models remains relatively unexplored.

Motivated by these developments, the present work investigates the dynamical behavior of a Hindmarsh--Rose neuron model augmented by magnetic flux under conditions supporting a blue-sky catastrophe. We focus on how electromagnetic coupling reshapes the intrinsic bifurcation structure of the neuron and drives transitions among periodic, bursting, and chaotic regimes. Using a combination of bifurcation analysis, phase-space characterization, Poincar\'e sections, and Lyapunov exponent calculations, we systematically map the parameter regions associated with different firing modes. 

By elucidating the combined effects of magnetic field influence and blue-sky catastrophe on neuronal dynamics, this study contributes to a deeper understanding of nonequilibrium mechanisms governing brain activity~\cite{NartalloKaluarachchi2026}. Our results provide new insights into how electromagnetic perturbations can induce or suppress chaos in neuronal systems, with potential implications for both theoretical neuroscience and neuromodulation strategies.

\section{Model and Results}
\subsection{Neuron model with blue-sky catastrophe and magnetic field}

We consider a single Hindmarsh--Rose (HR) neuron model exhibiting a blue-sky catastrophe, which is known to produce abrupt transitions between spiking and bursting dynamics~\cite{Bashkirtseva2019}. The governing equations are
\begin{align}
\frac{dx}{dt} &= y - a x^{3} + b x^{2} + I_{\text{ext}} - z, \label{eq:hrx} \\
\frac{dy}{dt} &= c - d x^{2} - y, \label{eq:hry} \\
\frac{dz}{dt} &= r\left[s(x-x_{0}) - z - \frac{\eta}{(z-z_{0})^{2} + \rho}\right], \label{eq:hrz}
\end{align}
where $x$ denotes the membrane potential, $y$ is a fast recovery variable associated with ionic activation, and $z$ represents a slow adaptation current governing bursting behavior. The control parameter $I_{\text{ext}}$ corresponds to an externally applied bias current. Parameter values are chosen as in Ref.~\cite{Bashkirtseva2019} to ensure the presence of a blue-sky catastrophe.

To investigate the influence of electromagnetic effects on neuronal dynamics, we extend the HR model by introducing a magnetic flux variable $\phi$, following earlier studies on electromagnetic induction in neuron models~\cite{LvMa2016,ThottilIgnatius2019}. The magnetic-field-coupled model is given by
\begin{align}
\frac{dx}{dt} &= y - a x^{3} + b x^{2} + I_{\text{ext}} - z - k_{1} W(\phi)x, \label{eq:mx} \\
\frac{dy}{dt} &= c - d x^{2} - y, \label{eq:my} \\
\frac{dz}{dt} &= r\left[s(x-x_{0}) - z - \frac{\eta}{(z-z_{0})^{2} + \rho}\right], \label{eq:mz} \\
\frac{d\phi}{dt} &= kx - k_{2}\phi + \phi_{\text{ext}}, \label{eq:phi}
\end{align}
where $\phi$ denotes the magnetic flux across the neuronal membrane and introduces an additional feedback channel coupling electromagnetic effects to the membrane potential.

The nonlinear interaction between magnetic flux and membrane voltage is described by the coupling function $W(\phi)$. In this work, we consider two representative functional forms,
\begin{align}
W(\phi) &= -\tanh(\phi), \label{eq:wtanh} \\
W(\phi) &= \alpha + 3\beta \phi^{2}, \label{eq:wpoly}
\end{align}
which enable us to assess the robustness of the observed dynamical behavior against different types of electromagnetic feedback.

The hyperbolic tangent form [Eq.~(\ref{eq:wtanh})] provides a bounded and saturating coupling, ensuring finite feedback strength for large magnetic flux and modeling saturation effects typical of electromagnetic induction. In contrast, the polynomial form [Eq.~(\ref{eq:wpoly})] yields an unbounded but smooth nonlinear modulation, allowing stronger amplification of electromagnetic effects. Although these two coupling mechanisms differ quantitatively, both generate qualitatively similar dynamical scenarios, including bursting transitions and chaos, indicating that the observed phenomena are not artifacts of a specific functional form.

External electromagnetic perturbations are incorporated through the term $\phi_{\text{ext}}=\varepsilon\xi(t)$, where $\xi(t)$ denotes a stochastic process and $\varepsilon$ controls the noise intensity. In the present work, we restrict our analysis to the deterministic limit by setting $\varepsilon=0$.

\subsection{Model parameters}

Unless stated otherwise, all parameters are fixed throughout the study and are chosen consistently with Refs.~\cite{Bashkirtseva2019,LvMa2016,ThottilIgnatius2019}. These values ensure the coexistence of slow--fast neuronal dynamics and the presence of a blue-sky catastrophe, allowing a systematic exploration of how magnetic-field-induced feedback modifies the bifurcation structure, firing patterns, and chaotic dynamics of the system.

For completeness and reproducibility, all parameter values and their physical interpretations are summarized in Table~\ref{tab:parameters}.

\begin{table}[]
\centering
\caption{Parameter values used in the Hindmarsh--Rose neuron model with magnetic flux.}
\label{tab:parameters}
\begin{tabular}{c c c}
\hline\hline
Parameter & Description & Value \\
\hline
$a$ & Nonlinear membrane parameter & $1.0$ \\
$b$ & Nonlinear membrane parameter & $3.0$ \\
$c$ & Recovery variable parameter & $1.0$ \\
$d$ & Recovery variable parameter & $5.0$ \\
$s$ & Slow adaptation gain & $4.0$ \\
$r$ & Slow timescale parameter & $0.006$ \\
$x_0$ & Reference membrane potential & $-1.6$ \\
$z_0$ & Reference adaptation current & $0.9$ \\
$\eta$ & Blue-sky catastrophe strength & $0.1$ \\
$\rho$ & Regularization parameter & $0.02$ \\
$k_1$ & Magnetic coupling strength & $0.95$ \\
$k_2$ & Magnetic damping coefficient & $0.5$ \\
$k$ & Flux--voltage coupling & $0.9$ \\
$\alpha$ & Baseline magnetic contribution (in $W$) & $0.01$ \\
$\beta$ & Quadratic magnetic nonlinearity (in $W$) & $0.02$ \\
$\varepsilon$ & Noise intensity & $0$ \\
\hline\hline
\end{tabular}
\end{table}

\subsection{Numerical methods}

The dynamical equations are integrated numerically using a standard fourth-order Runge--Kutta (RK4) scheme with a fixed time step. For all simulations reported in this work, the integration time step is chosen as $\Delta t = 0.01$, which provides an accurate resolution of both fast spiking dynamics and slow bursting timescales. Convergence tests with smaller time steps confirm that the qualitative and quantitative features of the dynamics remain unchanged.

For each value of the control parameter $I_{\mathrm{ext}}$, the system is evolved for a sufficiently long transient time to eliminate dependence on initial conditions. Data used for analysis are collected only after discarding the transient portion of the trajectory. Initial conditions are fixed as $(x,y,z,\phi) = (0.1,0,0,0)$ unless stated otherwise.

Bifurcation diagrams are constructed by recording successive maxima of the membrane potential $x(t)$ after transients. Chaotic and periodic regimes are further characterized using phase-space projections and Poincar\'e sections. The largest Lyapunov exponent is computed using the Wolf--Swift--Swinney--Vastano algorithm~\cite{Wolf1985}, which estimates exponential divergence of nearby trajectories through repeated renormalization of infinitesimal perturbations. Finite-time effects are minimized by long integration times and appropriate renormalization thresholds.

\subsection{Bifurcation structure under polynomial magnetic coupling}

\begin{figure}
    \centering
    \includegraphics[width=0.95\linewidth]{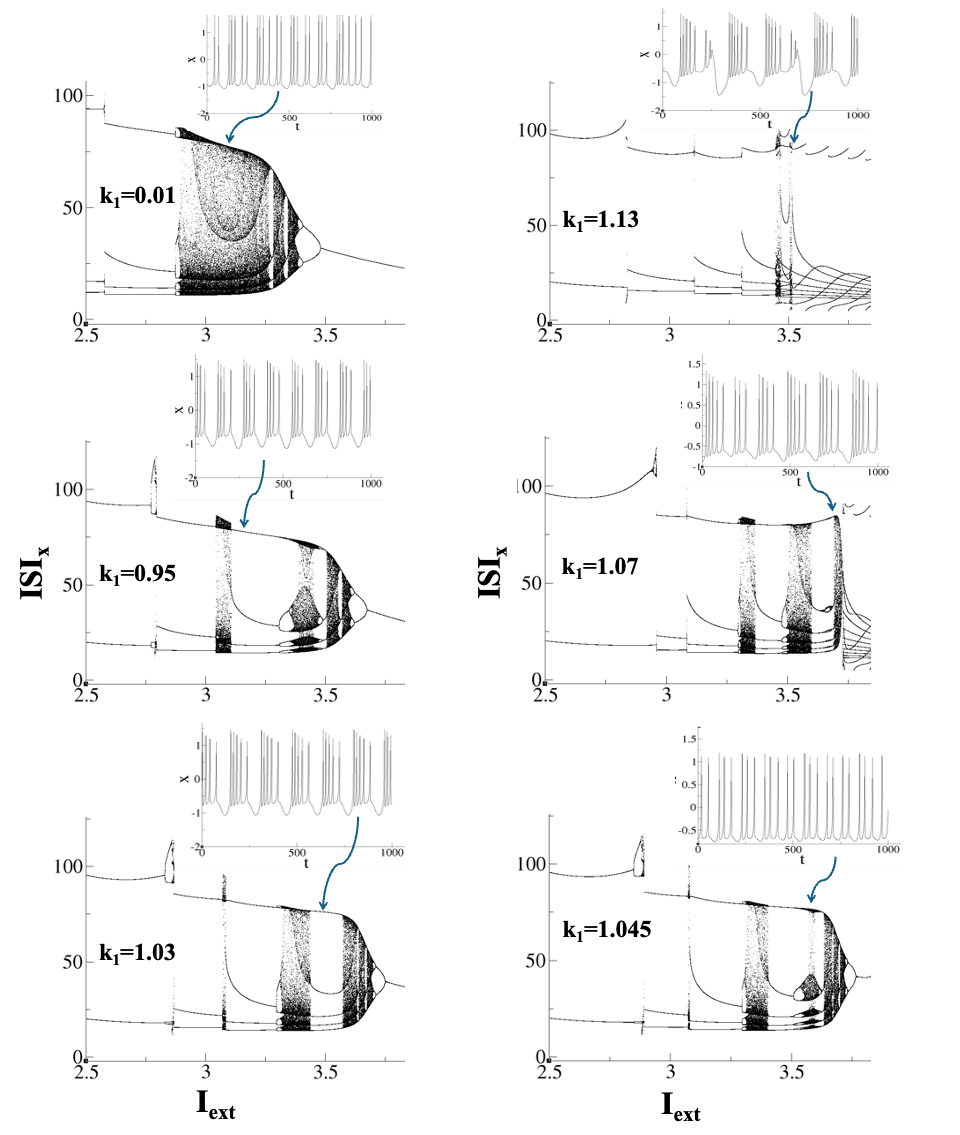}
    \caption{ISI bifurcation diagrams of the Hindmarsh–Rose neuron model with blue-sky catastrophe under polynomial magnetic coupling 
$W(\phi)=\alpha+3\beta\phi^{2}$
for different values of the magnetic coupling strength $k_1$. The external current $I_\mathrm{ext}$
 is used as the bifurcation parameter. Insets show representative membrane potential time series corresponding to selected parameter values (arrows). Increasing magnetic coupling induces fragmentation of periodic branches, intermittent dynamics, and chaotic bursting, followed by suppression of complexity at strong coupling.}
    \label{fig:1}
\end{figure}

Figure~\ref{fig:1} illustrates the interspike interval (ISI) bifurcation diagrams of the Hindmarsh--Rose neuron model with blue-sky catastrophe under the polynomial magnetic coupling
\[
W(\phi)=\alpha+3\beta\phi^{2},
\]
for different values of the magnetic coupling strength $k_1$. The bifurcation parameter is the external current $I_{\mathrm{ext}}$, while the ISI is computed from successive maxima of the membrane potential after discarding transients. Representative time series corresponding to selected parameter values are shown as insets in each panel.

For weak magnetic coupling ($k_1=0.01$), the bifurcation structure closely resembles that of the uncoupled system. A clear blue-sky catastrophe is observed near $I_{\mathrm{ext}}\approx 3.0$, marking an abrupt transition from regular spiking to bursting dynamics. The bursting regime is characterized by a dense distribution of ISI values, indicative of complex spike clustering and strong slow--fast interaction. In this regime, the magnetic feedback acts only as a weak perturbation and does not qualitatively alter the intrinsic neuronal dynamics.

As the magnetic coupling strength is increased ($k_1=0.95$ and $k_1=1.03$), significant deformation of the bifurcation structure becomes evident. The bursting region broadens in parameter space, and the ISI distribution becomes increasingly fragmented, signaling the onset of multistability and intermittent dynamics. In particular, narrow windows of periodic spiking emerge within the bursting regime, coexisting with irregular firing patterns. The insets reveal mixed-mode oscillations, where episodes of regular spiking alternate with irregular bursts, reflecting enhanced nonlinear feedback from the magnetic flux.

For intermediate coupling strengths ($k_1=1.045$ and $k_1=1.07$), the system exhibits pronounced chaotic bursting. The ISI bifurcation diagrams display highly scattered point clouds over a wide range of $I_{\mathrm{ext}}$, with the disappearance of well-defined periodic branches. These features are consistent with strong sensitivity to initial conditions and indicate a transition to chaos driven by the unbounded nonlinear magnetic feedback. The corresponding time series show irregular burst durations and variable interspike intervals, confirming the loss of temporal regularity.

At sufficiently strong coupling ($k_1=1.13$), the bifurcation structure undergoes a qualitative simplification. Large portions of the chaotic bursting region collapse into narrow bands, and the ISI distribution becomes sparse. This suggests that excessive magnetic feedback suppresses the slow adaptation dynamics responsible for bursting, effectively stabilizing simpler firing patterns. Such behavior highlights the nonmonotonic role of magnetic coupling: while moderate coupling enhances complexity and chaos, strong coupling can inhibit bursting and reduce dynamical richness.

Overall, Fig.~\ref{fig:1} demonstrates that the polynomial magnetic coupling profoundly reshapes the bifurcation landscape of the Hindmarsh--Rose neuron model. The blue-sky catastrophe persists across a wide range of coupling strengths, but its manifestation is strongly modulated by magnetic feedback. The emergence of multistability, intermittent dynamics, and chaotic bursting underscores the ability of electromagnetic effects to regulate neuronal firing patterns and induce complex transitions beyond those present in the uncoupled system.

 \subsection{Magnetic-field-induced modulation of bifurcation structure}

\begin{figure}
    \centering
    \includegraphics[width=0.95\linewidth]{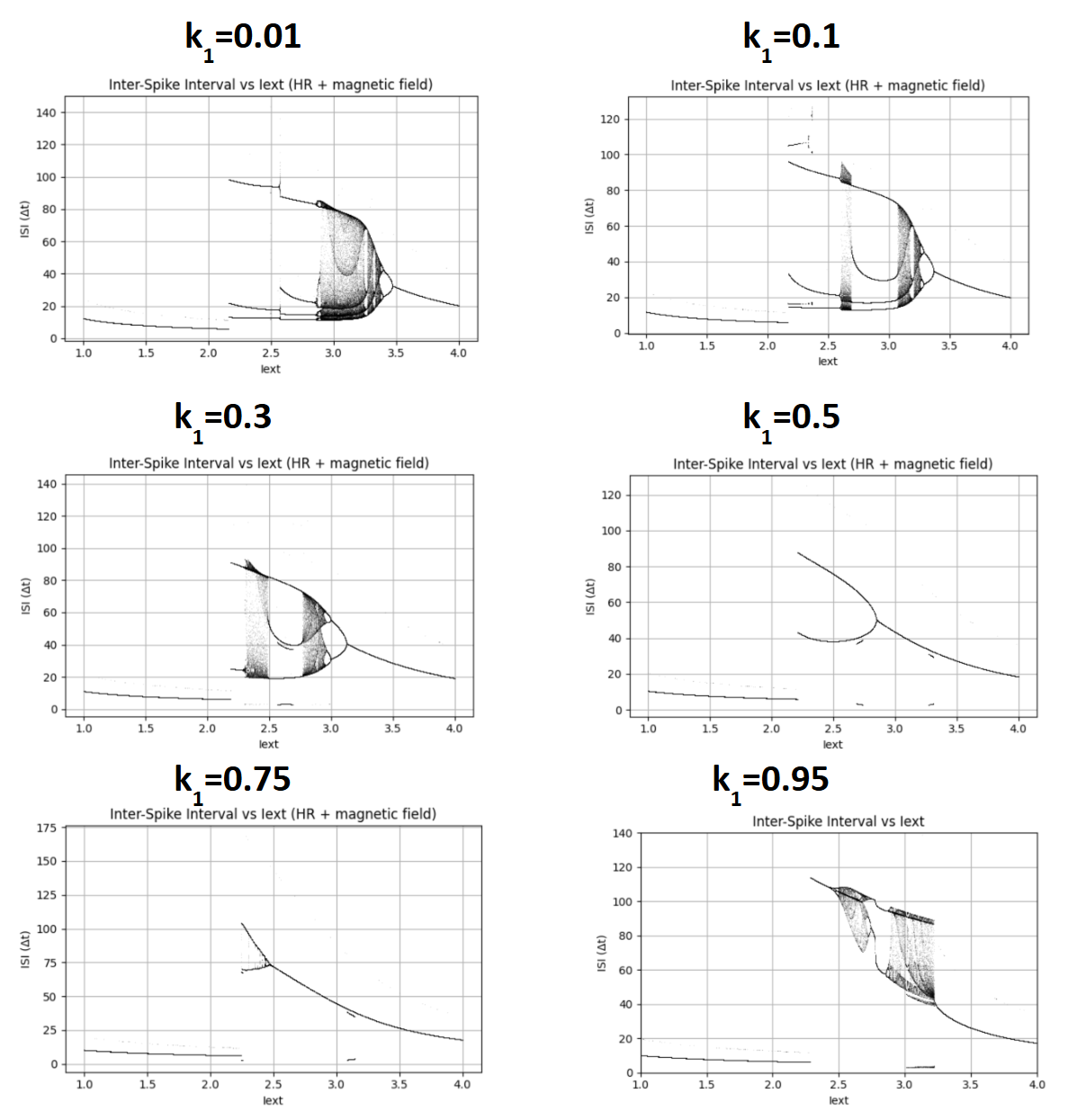}
    \caption{ISI bifurcation diagrams of the Hindmarsh–Rose neuron model with blue-sky catastrophe under polynomial magnetic coupling 
$W(\phi)=-\tanh(\phi)$
for different values of the magnetic coupling strength $k_1$. The external current $I_\mathrm{ext}$
 is used as the bifurcation parameter. Weak coupling preserves the intrinsic blue-sky catastrophe and bursting dynamics, moderate coupling stabilizes periodic spiking, while strong coupling reintroduces irregular and chaotic firing.}
    \label{fig:2}
\end{figure}

Figure~\ref{fig:2} presents the interspike interval (ISI) bifurcation diagrams of the Hindmarsh--Rose neuron model with blue-sky catastrophe under the polynomial magnetic coupling
\[
W(\phi)=-\tanh(\phi),
\]
for different values of the magnetic coupling strength $k_1$. The bifurcation parameter is the external current $I_{\mathrm{ext}}$, while the ISI is computed from successive maxima of the membrane potential after discarding transients. Representative time series corresponding to selected parameter values are shown as insets in each panel.

For weak magnetic coupling ($k_1=0.01$), the bifurcation diagram closely resembles that of the uncoupled system. A pronounced blue-sky catastrophe is observed near $I_{\mathrm{ext}}\approx 3.0$, marking an abrupt transition from regular spiking to bursting dynamics. The bursting regime is characterized by a dense cloud of ISI values, reflecting complex spike clustering governed primarily by the intrinsic slow--fast dynamics of the neuron. In this regime, the magnetic field acts as a weak perturbation and does not qualitatively modify the firing behavior.

As the coupling strength is increased ($k_1=0.1$ and $k_1=0.3$), noticeable deformation of the bifurcation structure occurs. The bursting region narrows, and the ISI distribution becomes less scattered, indicating partial suppression of complex bursting dynamics. Periodic spiking branches become more prominent, and several irregular ISI clusters disappear. This suggests that moderate magnetic feedback stabilizes the neuronal dynamics by attenuating the slow adaptation mechanisms responsible for bursting.

For intermediate values of the coupling strength ($k_1=0.5$ and $k_1=0.75$), the bifurcation structure simplifies further. The ISI diagram is dominated by smooth, well-defined branches corresponding to periodic spiking, while bursting and chaotic regimes are largely suppressed. The disappearance of dense ISI clouds indicates a loss of multistability and a reduction in dynamical complexity. In this parameter range, magnetic coupling effectively regulates the neuronal activity, favoring regular firing patterns over irregular bursting.

Interestingly, upon further increasing the coupling strength ($k_1=0.95$), complex dynamics re-emerge. The ISI bifurcation diagram again exhibits scattered point distributions over a broad range of $I_{\mathrm{ext}}$, signaling the onset of chaotic bursting. This nonmonotonic dependence on $k_1$ demonstrates that strong magnetic feedback can destabilize previously regular firing states and reintroduce irregular dynamics through enhanced nonlinear coupling between the magnetic flux and membrane potential.

Overall, Fig.~\ref{fig:2} reveals that magnetic-field-induced feedback plays a dual role in shaping neuronal dynamics. While moderate coupling suppresses bursting and chaos by stabilizing periodic spiking, stronger coupling can reintroduce complexity and chaotic firing. The persistence of the blue-sky catastrophe across a wide range of coupling strengths underscores its robustness, while the strong modulation of ISI patterns highlights the critical role of electromagnetic effects in regulating neuronal excitability and bifurcation structure.

\subsection{Strong magnetic coupling and emergence of structured irregular dynamics}

\begin{figure}[]
\centering
\includegraphics[width=0.95\linewidth]{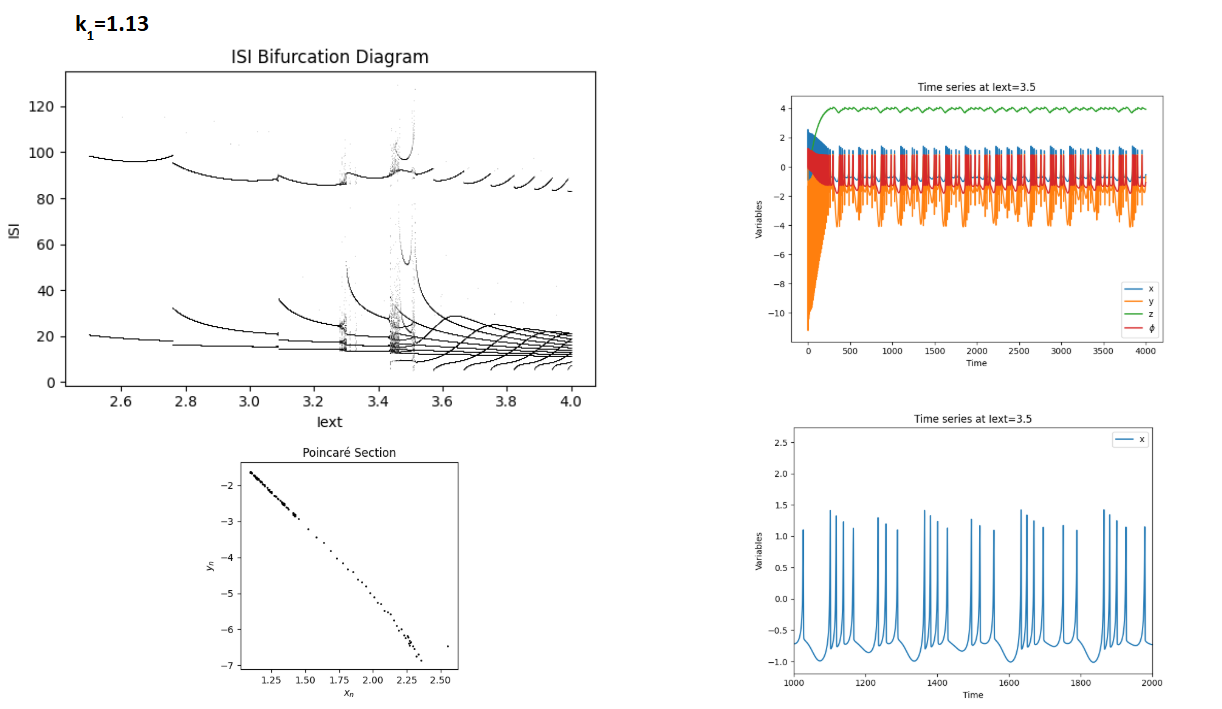}
\caption{
Dynamics of the Hindmarsh--Rose neuron model with magnetic-field coupling at $k_1 = 1.13$. 
(a) Interspike interval (ISI) bifurcation diagram as a function of the external current $I_{\mathrm{ext}}$, showing banded and irregular firing regimes.
(b) Time series of the state variables $x$, $y$, $z$, and $\phi$ at $I_{\mathrm{ext}} = 3.5$, illustrating amplitude-modulated spiking and slow adaptation dynamics.
(c) Poincar\'e section constructed from successive intersections of the trajectory, revealing a continuous curved set characteristic of quasiperiodic or weakly chaotic motion.
(d) Zoomed time series of the membrane potential $x(t)$ highlighting clustered spiking and variable interspike intervals.
}
\label{fig:k113}
\end{figure}

Figure~\ref{fig:k113} presents a comprehensive characterization of the neuronal dynamics at a strong magnetic coupling strength, $k_1 = 1.13$. The figure combines the interspike interval (ISI) bifurcation diagram, representative time series, and a Poincar\'e section, thereby providing a global and local perspective on the system’s behavior.

The ISI bifurcation diagram [Fig.~\ref{fig:k113}(a)] reveals a markedly structured yet irregular firing regime over a wide range of the external current $I_{\mathrm{ext}}$. In contrast to the dense chaotic clouds observed at intermediate coupling strengths, the ISI values here organize into narrow, folded bands interspersed with sparse point distributions. This indicates the coexistence of multiple firing modes with varying degrees of regularity, suggesting that strong magnetic feedback constrains the phase space while preserving sensitivity to parameter variations.

Time series of the state variables at a representative value $I_{\mathrm{ext}} = 3.5$ are shown in Fig.~\ref{fig:k113}(b). The membrane potential $x(t)$ exhibits repeated spiking with pronounced amplitude modulation, while the recovery and adaptation variables ($y$ and $z$) display slower oscillations synchronized with spike clusters. The magnetic flux variable $\phi(t)$ evolves on a comparable timescale and remains dynamically active, confirming its nontrivial role in shaping the observed firing patterns. Notably, the spiking activity is neither strictly periodic nor fully irregular, reflecting a structured form of irregular dynamics.

Further insight is provided by the Poincar\'e section shown in Fig.~\ref{fig:k113}(c). Instead of collapsing to a single point, which would indicate a stable limit cycle, the intersection points form a continuous, curved set. This geometry is characteristic of quasiperiodic or weakly chaotic motion and is consistent with the banded ISI structure observed in the bifurcation diagram. The absence of a fully scattered fractal set suggests that, although strong magnetic coupling reintroduces complexity, it also imposes constraints that limit the extent of chaotic wandering in phase space.

The zoomed time series of the membrane potential [Fig.~\ref{fig:k113}(d)] further illustrates the coexistence of regular spike timing with slow modulation, revealing clustered spikes separated by variable interspike intervals. Such behavior is indicative of mixed-mode oscillations, arising from the interplay between fast membrane dynamics, slow adaptation, and magnetic-field-induced feedback.

Overall, the results at $k_1 = 1.13$ demonstrate that strong magnetic coupling does not simply suppress or enhance chaos monotonically. Instead, it gives rise to a structured irregular regime in which complex firing patterns coexist with constrained phase-space geometry. This highlights the subtle and nontrivial role of electromagnetic effects in regulating neuronal excitability and dynamical complexity in systems exhibiting a blue-sky catastrophe.

\subsection{Chaotic bursting induced by intermediate magnetic coupling}

\begin{figure}[]
\centering
\includegraphics[width=0.95\linewidth]{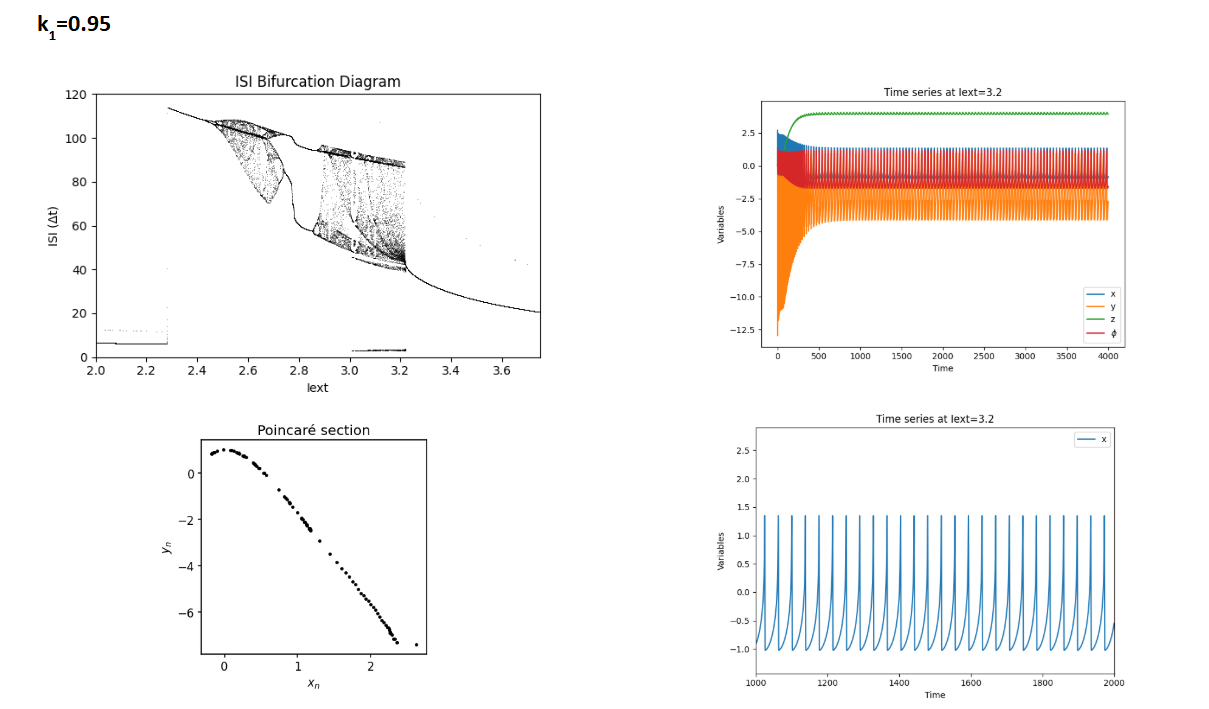}
\caption{
Dynamics of the Hindmarsh--Rose neuron model with magnetic-field coupling at $k_1 = 0.95$.
(a) Interspike interval (ISI) bifurcation diagram as a function of the external current $I_{\mathrm{ext}}$, showing dense and fragmented ISI distributions characteristic of chaotic bursting.
(b) Time series of the state variables $x$, $y$, $z$, and $\phi$ at $I_{\mathrm{ext}} = 3.2$, illustrating irregular bursting and slow modulation.
(c) Poincar\'e section constructed from successive intersections of the trajectory, revealing a scattered set associated with a chaotic attractor.
(d) Zoomed time series of the membrane potential $x(t)$ highlighting irregular interspike intervals within bursts.
}
\label{fig:k095}
\end{figure}

Figure~\ref{fig:k095} summarizes the dynamical behavior of the Hindmarsh--Rose neuron model with magnetic-field coupling at an intermediate coupling strength, $k_1 = 0.95$. The figure combines the ISI bifurcation diagram, time-series analysis, and a Poincar\'e section to characterize the nature of the firing dynamics.

The ISI bifurcation diagram shown in Fig.~\ref{fig:k095}(a) reveals a broad parameter region in which dense and irregular ISI distributions coexist with remnants of smooth branches. As the external current $I_{\mathrm{ext}}$ is varied, periodic spiking solutions undergo successive destabilization, giving rise to fragmented bands and scattered point clouds. This structure is indicative of chaotic bursting interspersed with narrow periodic windows, reflecting strong sensitivity to parameter variations.

Representative time series at $I_{\mathrm{ext}} = 3.2$ are shown in Fig.~\ref{fig:k095}(b). The membrane potential $x(t)$ exhibits bursting activity with irregular burst durations and variable interspike intervals. The recovery variable $y(t)$ and the slow adaptation variable $z(t)$ evolve on distinct timescales and display complex modulation synchronized with spike clusters. The magnetic flux variable $\phi(t)$ remains dynamically active and contributes to the modulation of bursting patterns, confirming the nontrivial role of electromagnetic feedback at this coupling strength.

The Poincar\'e section displayed in Fig.~\ref{fig:k095}(c) provides further evidence of chaotic dynamics. Instead of collapsing onto a single point or a smooth closed curve, the section forms a scattered, elongated set, characteristic of a strange attractor. This geometry confirms the presence of deterministic chaos in the bursting regime observed at $k_1 = 0.95$.

Figure~\ref{fig:k095}(d) shows a zoomed time series of the membrane potential $x(t)$, highlighting the irregular spacing between successive spikes within bursts. The absence of strict periodicity, combined with the scattered Poincar\'e points and dense ISI distributions, provides consistent evidence for chaotic bursting dynamics.

Taken together, these results demonstrate that intermediate magnetic coupling strongly enhances dynamical complexity, destabilizing regular spiking and promoting chaotic bursting through nonlinear interaction between membrane dynamics, slow adaptation, and magnetic flux feedback.

\subsection{Lyapunov characterization of bifurcation-induced complexity}

\begin{figure*}[]
\centering
\includegraphics[width=0.95\textwidth]{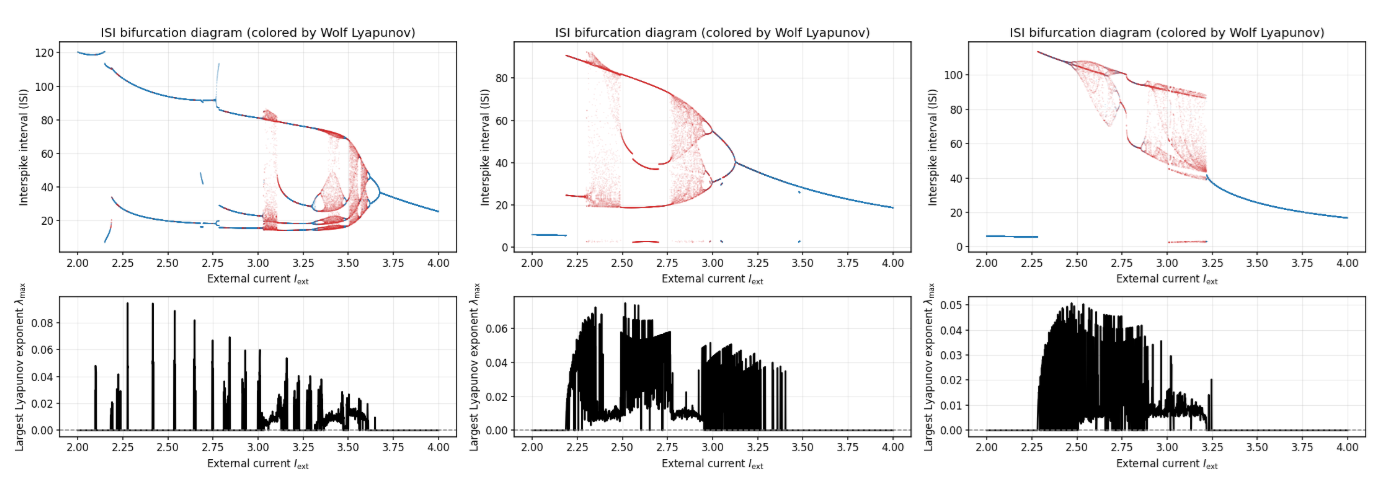}
\caption{
Correlation between ISI bifurcation structure and Lyapunov instability in the Hindmarsh--Rose neuron model with magnetic-field coupling.
Top row: Interspike interval (ISI) bifurcation diagrams as a function of the external current $I_{\mathrm{ext}}$, with points colored according to the sign of the largest Lyapunov exponent computed using the Wolf method (blue: $\lambda_{\max} \le 0$, red: $\lambda_{\max} > 0$).
Bottom row: Corresponding largest Lyapunov exponent $\lambda_{\max}$ evaluated over the same parameter range.
From left to right, the panels correspond to increasing magnetic coupling strength $k_1$.
Weak coupling yields predominantly regular dynamics with isolated chaotic windows, intermediate coupling produces extended chaotic regimes, and strong coupling leads to structured irregular dynamics with reduced but persistent instability.
}
\label{fig:lyap_bif}
\end{figure*}

Figure~\ref{fig:lyap_bif} presents a combined view of the interspike interval (ISI) bifurcation diagrams and the corresponding largest Lyapunov exponent $\lambda_{\max}$, computed using the Wolf algorithm, for representative values of the magnetic coupling strength $k_1$. The upper panels show the ISI bifurcation diagrams as a function of the external current $I_{\mathrm{ext}}$, with each point color-coded according to the sign of the local Lyapunov exponent, while the lower panels display $\lambda_{\max}$ evaluated over the same parameter range.

For weak magnetic coupling [left column], the ISI bifurcation diagram is dominated by smooth branches associated with periodic spiking and bursting dynamics. These regions are consistently colored blue, indicating nonpositive Lyapunov exponents. Positive Lyapunov values appear only in narrow parameter windows, coinciding with localized regions of fragmented ISI structure. This confirms that, at weak coupling, chaotic dynamics is confined to isolated intervals embedded within largely regular firing regimes.

At intermediate coupling strength [middle column], the bifurcation structure becomes substantially more complex. Large portions of the ISI diagram are colored red, signaling positive Lyapunov exponents over extended ranges of $I_{\mathrm{ext}}$. These regions correspond to dense ISI clouds and fragmented branches, characteristic of chaotic bursting and intermittent dynamics. The Lyapunov spectrum exhibits sustained positive values, interspersed with sharp drops to near-zero levels, reflecting the coexistence of chaotic and nearly periodic solutions. This regime represents a maximal enhancement of dynamical complexity induced by magnetic-field feedback.

For stronger coupling [right column], the ISI bifurcation diagram shows a partial reorganization of firing patterns. While extended chaotic regions persist, as indicated by red coloring and positive Lyapunov exponents, the overall structure becomes more constrained. The Lyapunov spectrum displays a broad plateau of positive values followed by a gradual decline as $I_{\mathrm{ext}}$ increases, consistent with the emergence of structured irregular or weakly chaotic dynamics rather than fully developed chaos. This behavior suggests that strong magnetic coupling limits phase-space exploration while maintaining sensitivity to initial conditions.

Taken together, Fig.~\ref{fig:lyap_bif} establishes a direct correspondence between geometrical features of the ISI bifurcation diagrams and quantitative measures of dynamical instability. Periodic spiking and regular bursting are associated with nonpositive Lyapunov exponents, whereas fragmented ISI distributions and dense point clouds reliably coincide with positive $\lambda_{\max}$. The results demonstrate that magnetic-field-induced feedback not only reshapes the bifurcation landscape but also controls the onset, extent, and intensity of chaos in the Hindmarsh--Rose neuron model exhibiting a blue-sky catastrophe.

\section{Summary and discussion}

We have investigated the dynamical consequences of magnetic-field-induced feedback in a Hindmarsh--Rose neuron model exhibiting a blue-sky catastrophe. By augmenting the classical model with a magnetic flux variable, we demonstrated that electromagnetic coupling profoundly reshapes the bifurcation structure, firing patterns, and instability properties of the system.

Our analysis reveals that magnetic feedback plays a nonmonotonic role in neuronal dynamics. Weak coupling preserves the intrinsic blue-sky catastrophe and regular spiking--bursting transitions of the uncoupled system. At intermediate coupling strengths, magnetic feedback destabilizes periodic solutions, giving rise to fragmented bifurcation structures, chaotic bursting, and extended regions of positive Lyapunov exponents. Stronger coupling constrains phase-space exploration and leads to structured irregular dynamics, characterized by banded ISI distributions, quasiperiodic or weakly chaotic motion, and reduced instability. These regimes are consistently supported by time-series analysis and Poincar\'e sections.

By comparing bounded and polynomial forms of the magnetic coupling function, we showed that the qualitative dynamical scenarios, including the persistence of the blue-sky catastrophe and the emergence of chaos, are robust with respect to the specific functional form of electromagnetic feedback. However, the strength and extent of chaotic regimes depend sensitively on the coupling intensity, highlighting the regulatory role of magnetic effects.

The combined use of ISI bifurcation diagrams and Wolf Lyapunov exponents establishes a clear correspondence between geometric features of firing patterns and quantitative measures of instability. Periodic and quasiperiodic regimes are associated with nonpositive Lyapunov exponents, whereas fragmented ISI structures reliably coincide with positive Lyapunov values, confirming the deterministic origin of the observed complexity.

Overall, our results demonstrate that electromagnetic interactions can serve as an effective control mechanism for neuronal excitability and chaos in systems undergoing blue-sky catastrophes. These findings provide a dynamical foundation for understanding how external fields and intrinsic electromagnetic processes may regulate neuronal firing patterns and transitions between regular and chaotic activity.

\section*{Acknowledgements}
RPY acknowledges Priya Gupta and Sharukh Tyagi for critical inputs and discussions.

\bibliography{apssamp}

\end{document}